\documentclass[prl,twocolumn,showpacs,floatfix,citeautoscript]{revtex4}
\usepackage{graphicx}
\usepackage{amsmath}

\begin{document}

\title{Flux Qubits and Readout Device with Two Independent Flux Lines}

\author{B.L.T. Plourde}
\author{T.L. Robertson}
\author{P.A. Reichardt}
\author{T. Hime}
\author{S. Linzen}
\author{C.-E. Wu}
\author{John Clarke}
\affiliation{Department of Physics, University of California, 
Berkeley, CA 94720-7300}

\date{\today}

\pacs{03.67.Lx, 85.25.Cp, 85.25.Dq}

\begin{abstract}
We report measurements on two superconducting flux qubits coupled to a
readout Superconducting QUantum Interference Device (SQUID). Two
on-chip flux bias lines allow independent flux control of any two of
the three elements, as illustrated by a two-dimensional qubit flux
map. The application of microwaves yields a frequency-flux dispersion
curve for 1- and 2-photon driving of the single-qubit excited
state, and coherent manipulation of the single-qubit state results in
Rabi oscillations and Ramsey fringes. This architecture should be
scalable to many qubits and SQUIDs on a single chip.
\end{abstract}

\maketitle

Superconducting quantum bits (qubits) based on charge
\cite{Nakamura99,Vion02}, magnetic flux \cite{Friedman00,vanderWal00},
and phase difference across a Josephson junction
  \cite{Ramos01,Martinis02} are attractive candidates for the basis
of a quantum computer because of their inherent scalability using
established thin-film fabrication techniques.  Advantages of the flux
qubit include its immunity to the ubiquitous charge noise in the
substrate and that it can be configured with no direct electrical
connections.  One type of flux qubit consists of a superconducting
loop interrupted by three Josephson junctions with critical currents
$I_0$, $I_0$, and $\alpha I_0$ ($\alpha < 1$) \cite{Mooij99}. When the
applied flux bias $\Phi_Q$ is at a degeneracy point $(n+1/2)\Phi_0$
($n$ is an integer such that $| \Phi_Q-n \Phi_0| \leq \Phi_0/2$;
$\Phi_0\equiv h/2e$ is the flux quantum), a screening supercurrent
$J_Q$ can flow in either direction around the loop.  The ground and
first excited states of the qubit correspond to symmetric and
antisymmetric superpositions of the two current states and are
separated by an energy $\Delta$.  Here, $\Delta/h$ is the tunnel
frequency between the current states, typically a few GHz.  When
$\Phi_Q$ is away from a degeneracy point, the energy difference
between the two superposed states is $\nu = (\Delta^2 +
\epsilon^2)^{1/2}$, where $\epsilon = 2J_Q [\Phi_Q - (n +1/2)
  \Phi_0]$.  The state of the qubit is measured by coupling the
screening flux generated by $J_Q$ to a hysteretic dc superconducting
quantum interference device (SQUID). This flux determines the bias
current at which the SQUID switches out of the zero-voltage state.

In addition to the development of scalable interqubit couplings
\cite{Plourde04}, a prerequisite for scaling to a system of many
qubits is that the attendant readout, filtering, and bias circuitry
also scale.  A particular challenge is that the flux bias must be
settable for each element individually. This mandates the use of
on-chip flux-bias lines in an arrangement that enables one to apply a
combination of currents to address any given qubit or SQUID while
maintaining all other flux biases at constant values.  Furthermore,
the bias currents required to change the flux over (say) $\pm 1\Phi_0$
should not be so large that it becomes impractical to deliver them to
a chip cooled to millikelvin temperatures.  This requirement
establishes minimum self-inductances of the qubit and readout SQUID
that are substantially larger than values used previously in
3-junction qubits, which have relied on external coils to generate
large magnetic fields \cite{vanderWal00, Chiorescu03,Chiorescu04}. At
the same time, the mutual inductance between the on-chip flux lines
and the qubit must be sufficiently small for the noise generated by
the circuitry supplying the flux bias current not to be the limiting
source of decoherence \cite{Makhlin01}.

In this Letter, we report measurements on two qubits and a readout
SQUID that meet these criteria.  We illustrate the orthogonalization
of the applied fluxes by means of a two-dimensional flux map.  We
report spectroscopy on one of the two qubits that matches the expected
dispersion for a flux qubit.  In addition, we observe spurious
resonances, some of which may be related to defects in the tunnel
barriers of the qubit.  We perform coherent manipulation of the
single-qubit state resulting in Rabi oscillations and Ramsey fringes.
In prior measurements of coherent oscillations
\cite{Chiorescu03,Chiorescu04}, the readout SQUID was connected
directly to the flux qubit and thus detected a combination of flux and
phase changes. In contrast, our SQUID is electrically isolated from
the qubit and detects only flux changes.

\begin{figure}
\includegraphics{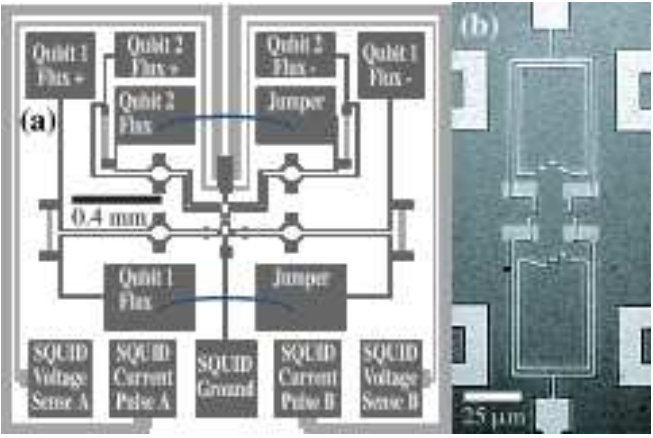}
\caption{ (a) Chip layout.  Dark gray represents Al traces, light gray
  AuCu traces.  Pads near upper edge of chip provide two independent
  flux lines; wirebonded Al jumpers couple left and right halves.
  Pads near lower edge of chip supply current pulses to the readout
  SQUID and sense any resulting voltage.  
  (b) Photograph of center region of completed device.  Segments of
  flux lines are visible to left and right of SQUID, which surrounds
  the two qubits.  
\label{fig:layout}}
\end{figure}

Figure 1 shows the device layout.  The readout SQUID has a
calculated inductance $L_S=358$ pH, and each of the two qubits it
encloses has a calculated inductance $L_Q=143$ pH.  The calculated
mutual inductance between each qubit and the SQUID is $61$ pH.  One
pair of series-connected flux bias lines is arranged near the top of
the SQUID and a second near the bottom; thus, flux in qubit 1 (2) is
supplied predominantly by the lower (upper) flux lines.  The mutual
inductance between each qubit and its associated flux lines was
designed to be $4$-$5$ pH, enabling us to apply $\sim 2\Phi_0$ with a
current of $1$ mA.  This criterion dictated the relatively large qubit
self-inductances compared with those in previous experiments 
\cite{vanderWal00, Chiorescu03, Saito04, Jena04}.

We fabricated the device on an oxidized Si substrate using
electron-beam lithography and double-angle evaporation to form the
Al-AlOx-Al tunnel junctions.  The Al lines for the qubit and SQUID
loops were $1$ $\mu$m wide, and those for the flux bias $10$ $\mu$m
wide.  Each SQUID junction was $175 \times 200$ nm$^2$ with a critical
current of $220$ nA.  For qubits 1 and 2, the larger junctions had
areas of $250 \times 250$ nm$^2$ and $180 \times 200$ nm$^2$,
approximate critical currents $I_0$ (scaled from SQUID junction areas)
of $390$ nA and $230$ nA, and $\alpha$-values (based on junction
areas) of $0.49$ and $0.68$, respectively.  The different junction
parameters were chosen to increase the probability of obtaining one
viable qubit; in fact, qubit 2 displayed good characteristics.  A $42$
nm-thick AuCu film deposited and patterned prior to the Al deposition
provided quasiparticle traps near the junctions \cite{Lang03}, $100$
$\Omega$ shunts on each of the four flux lines, and $500$ $\Omega$ and
$1275$ $\Omega$ series resistors on each end of the pulse lines and
sense lines, respectively.

To eliminate external magnetic field fluctuations we enclosed the chip
in a $6 \times 16 \times 22$ mm$^3$ cavity machined into a copper
block and plated with Pb.  This superconducting cavity stabilizes the
field, enabling us to acquire data for periods up to 48 hours.  A $1$
mm-diameter superconducting loop $\sim3$ mm above the chip supplied
microwave flux.  The sample holder was attached to the mixing chamber
of a dilution refrigerator at $50$ mK.  All electrical leads to the
experiment were heavily filtered at several different temperatures
with a combination of lumped circuit and copper powder low-pass
filters \cite{Martinis87}. Measurements of the state of a qubit as a
function of flux, microwave power and frequency were made by pulsing
the current in the SQUID and detecting whether or not it switched out
of the zero-voltage state by means of a low-noise, room-temperature
amplifier.  The flux bias currents were supplied by highly stable
potentiometers, controlled by a computer over a fiber-optic link
\cite{Linzen04}.

There are three applied fluxes that determine the state of the system:
the SQUID flux $\Phi_S$, and the qubit fluxes $\Phi_{Q1}$ and
$\Phi_{Q2}$.  Given the two flux lines, we can set any two fluxes
arbitrarily; the third is fully constrained.  This can be expressed
succinctly by the matrix equation
\begin{equation}
\begin{bmatrix}
\Phi_S&\Phi_{Q1}&\Phi_{Q2}
\end{bmatrix}
=
\begin{bmatrix}
M_{F1S}&M_{F2S}&\Phi_S^0\\
M_{F1Q1}&M_{F2Q1}&\Phi_{Q1}^0\\
M_{F1Q2}&M_{F2Q2}&\Phi_{Q2}^0\\
\end{bmatrix}
\begin{bmatrix}
I_1\\I_2\\1
\end{bmatrix},
\label{eq:mutuals}
\end{equation}
where the $M_{ij}$ are mutual inductances between the various flux
lines and loops, the $\Phi_j^0$ account for static background fields,
and the $I_i$ are currents in the flux lines.  

We turn now to our experimental results.  Figure 2(a) shows the
switching probability of the SQUID versus the amplitude of the current
pulses applied to it for two different values of 
$\Phi_{Q2}$ at constant $\Phi_S$.  
Because we hold
$\Phi_S$ constant, we can measure the displacement of the two curves
for constant sensitivity in the SQUID.  The measurement fidelity,
which is the difference between the switching probabilities, has a
maximum of about $60\%$.  
Figure 2(b) shows $I_s^{50\%}$ vs. $\Phi_S$; $I_s^{50\%}$ is the pulse
amplitude for which the switching probability is $50\%$.
The effects of the changing flux in the qubits are small on this
scale.  Figure 2(c) shows $I_s^{50\%}$ vs.  $\Phi_{Q1}$ for constant
$\Phi_S$; as $\Phi_{Q1}$ is varied, a flux of approximately the same
magnitude and opposite sign is applied to qubit 2.  
Thus, as $\Phi_{Q1}$ is increased at constant $\Phi_S$,
$I_s^{50\%}$ abruptly increases when qubit 1 flips state and decreases
when qubit 2 flips state.

\begin{figure}
\includegraphics{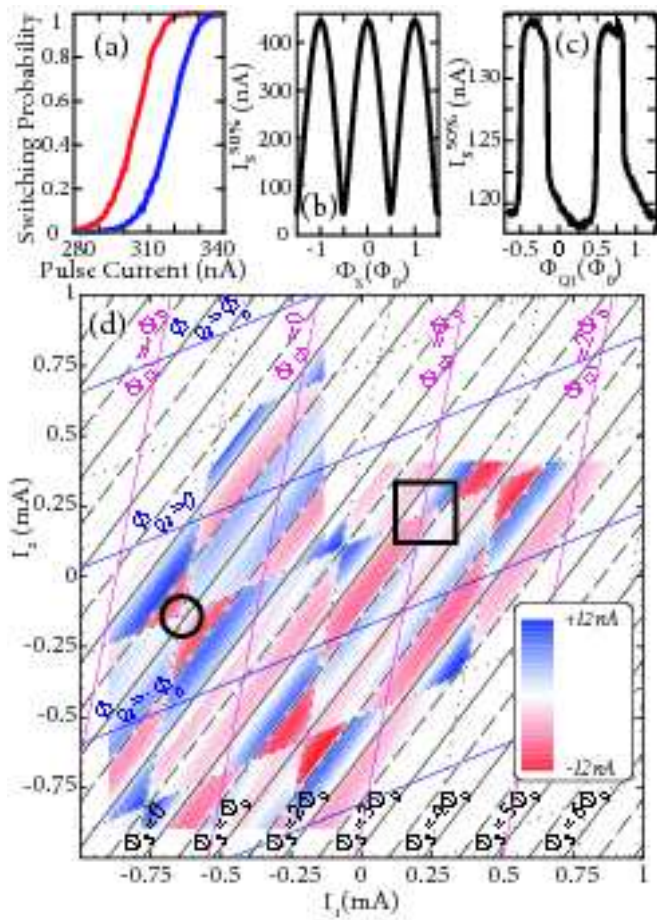}
\caption{(a) SQUID switching probability vs. amplitude of bias current
  pulse near qubit 2 transition.  The two curves represent the states
  corresponding to $\Phi_{Q2}=0.48 \Phi_0$ (red) and $\Phi_{Q2}=0.52
  \Phi_0$ (blue); $\Phi_S$ is held constant.  Each curve contains
  $100$ points averaged $8,000$ times.  (b) $I_s^{50\%}$ \emph{vs}.
  $\Phi_S$.  Each period of oscillation contains $\sim 5,000$ flux values,
  and each switching current is averaged $8,000$ times.  (c)
  Dependence of $I_s^{50\%}$ on $\Phi_{Q1}$ for constant $\Phi_S$.
  (d) Qubit flux map.
  \label{fig:flux-map}}
\end{figure}

To determine the parameters in Eq.~(\ref{eq:mutuals}), we sampled
$I_s^{50\%}$ at $\sim20,000$ different settings of $I_1$ and $I_2$.
We fit these data to a parametric model of the response of the
SQUID $I_s^{fit}(\Phi)$ to the total SQUID flux $\Phi_T$.  To describe the
SQUID modulation, we use the \emph{ad hoc} expression 
\begin{equation}
I_s^{fit}(\Phi)=\sum_{i=1}^{15} a_i \cos 2\pi i \frac{\Phi}{\Phi_0}
+\sum_{i=2}^{15} b_i \sin 2\pi i \frac{\Phi}{\Phi_0}  +d,
\end{equation}
where the $a_i$, $b_i$, and $d$ are fit parameters.  The total flux
coupled to the SQUID is well approximated by
\begin{equation}
\Phi_T=\Phi_S+ \Delta\Phi_{Q1} j_{Q1}(\Phi_{Q1})+\Delta\Phi_{Q2} j_{Q2}(\Phi_{Q2}),
\label{eq:total-flux}
\end{equation}
which neglects the self-screening of the SQUID. Here, the
$\Delta\Phi_{Qi}=J_{Qi} M_{QiS}$ are the 
amplitudes of qubit screening flux changes
near the degeneracy point, referred to the SQUID, 
and the qubit circulating currents are
described by dimensionless periodic functions of unit amplitude
\begin{equation}
j_{Qi}(\Phi)=\left\{\sin \left[2\pi g_{Qi} (\Phi-n\Phi_0)/\Phi_0\right]\right\}/\sin \pi g_{Qi},
\end{equation}
with fitting parameters $g_{Qi}$. 
The $j_{Qi}(\Phi)$ are discontinuous when $n$ increments.

Fitting the $I_s^{50\%}(I_1,I_2)$ data, we obtain the following
parameters for $I_s^{fit}(\Phi_T)$: $M_{F1Q1}=3.96$ pH,
$M_{F2Q1}=-0.77$ pH, $M_{F1Q2}=1.37$ pH, $M_{F2Q2}=-3.31$ pH,
$M_{F1S}=8.30$ pH, $M_{F2S}=-6.30$ pH, $\Delta\Phi_{Q1}=4.39$
m$\Phi_0$, $\Delta\Phi_{Q2}=4.04$ m$\Phi_0$, $g_{Q1}=0.64$,
$g_{Q2}=0.51$. This fit captures all the essential features of the
data, having a root-mean-square residual of $1.8$ nA, that is, less
than $0.5\%$ of the maximum value of $I_s^{50\%}$.

In Fig. 2(d), we plot $I_s^{50\%} - I_s^{fit}(\Phi_S)$ versus the
currents $I_1$ and $I_2$; lines of constant $\Phi_S$, $\Phi_{Q1}$, and
$\Phi_{Q2}$ are indicated. This qubit flux map displays only the flux
contributions of the two qubits.  For example, inside the square,
where $\Phi_{Q1} \approx \Phi_0$ and $\Phi_{Q2} \approx -\Phi_0/2$,
$I_s^{50\%}$ exhibits an abrupt step across the $\Phi_{Q2}$ line, as
$J_{Q2}$ changes direction discontinuously.  However, there is no step
across the $\Phi_{Q1}$ line, as $J_{Q1}$ is continuous.  Furthermore,
one can scan the switching current along a line where $\Phi_S$ is held
constant at an arbitrary value, to observe the effects of flux only on
the qubits.  Finally, special points where the two qubits are each at
a degeneracy point are indicated by the circle.

The remaining discussion of the experiments is concerned with the
effects of an applied microwave flux on qubit 2.  Photons of energy $h
f_m=\nu$ drive transitions between the two qubit states, producing
peaks and dips on the qubit transitions.  By measuring $I_s^{50\%}$ as
a function of $\Phi_{Q2}$ and $f_m$, we determined the dispersion
relation shown in Fig. 3. This measurement contains a total of $\sim
75,000$ points and took $48$ hours to acquire, thus demonstrating the
excellent flux stability in our system.  The dispersion is well
described by the hyperbolic relation
$\nu=(\epsilon^2+\Delta^2)^{1/2}$, with $\Delta/h=(3.99\pm 0.05)$ GHz
and $(1/h) d\epsilon/d\Phi_{Q2}=(896\pm 5)$ MHz/m$\Phi_0$. Part of the
spectrum corresponding to 2-photon transitions is also shown. 

\begin{figure}
\includegraphics{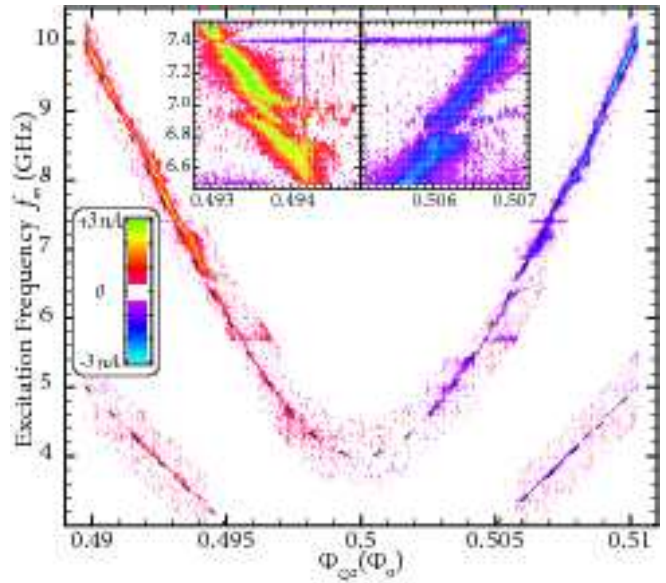}
\caption{Spectroscopy of qubit 2. 
  Enhancement and suppression of $I_s^{50\%}$ is shown as
  a function of $\Phi_{Q2}$ and $f_m$ relative to
  measurements in the absence of microwaves.  Dashed lines indicate
  fit to hyperbolic dispersion for 1- and 2-photon qubit
  excitations. The 2-photon fit is one-half the frequency of
  the 1-photon fit. Inset containing $\sim 23,000$ points is at higher
  resolution.
  \label{fig:spectroscopy}}
\end{figure}

Figure 3 shows two types of deviation from ideal behavior, more
evident in the inset which was measured $4$ weeks before the full
spectrum.  First, there are sharp suppressions of the critical current
which are independent of $\Phi_{Q2}$ and hence occur at particular
constant values of $f_m$, for example, at $7.4$ GHz, that we believe
arise from electromagnetic modes which couple to the qubit.
Second, we see disruptions
of the dispersion curve that are suggestive of coupling between the
qubit and other two-state systems.  One instance of this second
anomaly, originally centered near $7$ GHz in the inset and shifted to
$6.5$ GHz in the full spectrum, is remarkably similar to those
reported for phase qubits, and may be of the same origin
\cite{Simmonds04}.

\begin{figure}
\includegraphics{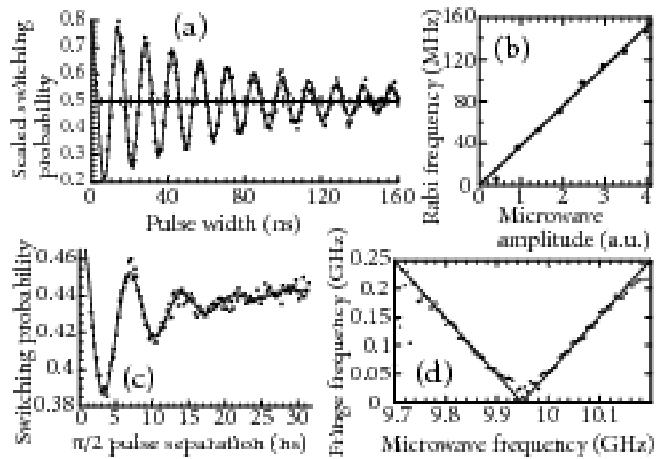}
\caption{ Coherent manipulation of qubit state.  (a) Rabi
  oscillations, scaled to measured SQUID fidelity, 
  as a function of width of $10.0$ GHz microwave pulses. 
  (b) Rabi frequency vs. $10.0$ GHz
  pulse amplitude; line is least squares fit to the data.  (c) Ramsey
  fringes for qubit splitting of $9.95$ GHz, microwave frequency of
  $10.095$ GHz. (d) Ramsey fringe frequency vs. microwave frequency.
  Lines with slopes $\pm 1$ are fits to data.
\label{fig:rabi}}
\end{figure}

We performed coherent manipulation of the qubit state by varying the
duration of a resonant microwave pulse for fixed frequency and
amplitude.  Upon applying fixed-amplitude bias current pulses, we
observed Rabi oscillations in the switching probability of the SQUID
as a function of the microwave pulse width at approximately 100
frequencies ranging from 4.37 to 13.86 GHz.  Generally speaking, the
amplitude of the oscillations is higher and their decay time is longer
for qubit flux bias points away from the spurious splittings shown in
the spectroscopy.  An example is shown in Fig. 4(a), where each of the
$195$ points was averaged $20,000$ times.  We reference the amplitude
of these oscillations to the qubit by scaling to the SQUID measurement
fidelity.  We observed Rabi oscillations with the longest decay time
when we operated at a bias current of $296$ nA, where the fidelity was
$14.4$\% for this particular value of $\Phi_S$ and the SQUID switched
infrequently [Fig. 2(a)].  We believe that this improvement is related
to noise currents in the SQUID loop which couple to the qubit through
$M_{Q2S}$ and are produced by the quasiparticles generated in the
SQUID loop during each switching event: a lower SQUID switching
probability results in fewer quasiparticles averaged over time
\cite{Reichardt04APS}.  Switching probability data in Fig.~4(a),
scaled to the measurement fidelity of the SQUID, fit well to a damped
sinusoid with a visibility of $63\%$ and a decay time $\tau_{Rabi}=78$
ns. Fits to similar measurements for different microwave pulse
amplitudes show that the Rabi frequency scales linearly with the
microwave amplitude [Fig. 4(b)], as expected for coherent driving
\cite{Rabi37}.

We measured the decay of a resonance peak and obtained the relaxation
time $\tau_R=281$ ns at $10.0$ GHz from an exponential fit.  We
measured the dephasing time $\tau_\phi$ from Ramsey fringes
\cite{Ramsey50}. We first applied a $\pi /2$ pulse to tip the qubit
state vector into the equatorial plane, where it dephased.  To
calibrate the $\pi /2$ pulses, we chose a microwave pulse width based
on a Rabi oscillation measurement made at the same operating point.
After a variable time delay $\tau$, we applied a second $\pi /2$ pulse
followed by a measurement of the SQUID switching probability for a
fixed bias current pulse amplitude. When we chose a microwave
frequency off resonance we observed damped oscillations of the SQUID
switching probability [Fig.  4(c)] with a fit decay time of the
oscillations $\tau_{\phi}=6.6$ ns.  Figure 4(d) shows the Ramsey
fringe frequency for $100$ values of the microwave frequency, together
with the fit to two lines with slopes of magnitude unity, as expected
for coherently driven Ramsey fringes.

In conclusion, we have demonstrated on-chip flux bias lines which
allow us to vary the flux applied to two of the three devices
independently. The mutual inductance between the flux lines and the
qubits is weak enough that the characteristic impedance of the flux
lines (say $50$ $\Omega$) should not limit our coherence times
\cite{Makhlin01}. Although in this Letter we have concentrated on the
quantum coherent properties of a single flux qubit, we note that the
two flux qubits in this design in principle could be coupled
controllably using the circulating currents in the dc SQUID
\cite{Plourde04}.  This system of qubits, SQUID and flux lines should
be readily scalable.

We thank M.H. Devoret, D. Esteve, C.J.P.M. Harmans, J.M. Martinis,
R. McDermott, J.E. Mooij, R.J. Schoelkopf, D. Vion, and F.K. Wilhelm
for helpful discussions.  This work was supported by the Air Force
Office of Scientific Research under Grant F49-620-02-1-0295, the Army
Research Office under Grant DAAD-19-02-1-0187, the National Science
Foundation under Grant EIA-020-5641, and the Advanced Research and
Development Activity.

\bibliography{i-f-qubits}

\end{document}